\documentclass[%12pt,         %
aps,                    %  American Physical Society
prd,                    %  Physical Review Letters
superscriptaddress,    %  Authors' addresses linked with superscripts
nofootinbib,            %  Does not treat footnotes as references
amsmath,               %
amssymb,               %
a4paper,                %  A4 paper format
floatfix]               %  Fixes float errors
{revtex4}               %  REVTEX 4 Package used

\usepackage{booktabs}
\usepackage{multirow}
\usepackage[colorlinks = true,
            linkcolor = blue,
            urlcolor  = blue,
            citecolor = blue,
            anchorcolor = blue]{hyperref}

\usepackage[pdftex]{graphics}
\usepackage{float}
\usepackage{amssymb}
\usepackage{graphics}
\usepackage{graphicx}
\usepackage{amsmath}
\usepackage{hyperref}
\usepackage{amsfonts}
\usepackage{subfigure}
\usepackage{xcolor}
\usepackage{cancel}
\usepackage{colordvi}

\newcommand{\fr}{\frac}

\newcommand{\la}{\lambda}
\newcommand{\La}{\Lambda}
\newcommand{\si}{\sigma}

\newcommand{\ga}{\gamma}

\newcommand{\ld}{\delta}

\newcommand{\be}{\begin{equation}}
\newcommand{\ee}{\end{equation}}
\newcommand{\beqa}{\begin{eqnarray}}
\newcommand{\eeqa}{\end{eqnarray}}
\newcommand{\bi}{\begin{itemize}}
\newcommand{\ei}{\end{itemize}}
\newcommand{\ben}{\begin{enumerate}}
\newcommand{\een}{\end{enumerate}}

\begin{document}

\title{f(R) Gravity in an Ellipsoidal Universe}

\author{Cemsinan Deliduman}
\email{cdeliduman@gmail.com}
\affiliation{Department of Physics, Mimar Sinan Fine Arts University, Bomonti 34380, \.{I}stanbul, Turkey}

\author{O\u{g}uzhan Ka\c{s}\i k\c{c}\i}
\email{kasikcio@itu.edu.tr}
\affiliation{Physics Engineering Department, \.{I}stanbul Technical University, Maslak 34469, \.{I}stanbul, Turkey}

\author{Vildan Kele\c{s} Tu\u{g}yano\u{g}lu}
\email[Corresponding author: ]{vildantugyanoglu@gmail.com}
\affiliation{Department of Physics, Mimar Sinan Fine Arts University, Bomonti 34380, \.{I}stanbul, Turkey}

\begin{abstract}
We propose a new model of cosmology based on an anisotropic background and a specific $f(R)$ theory of gravity. It is shown that field equations of $f(R)$ gravity in a Bianchi type I background give rise to a modified Friedmann equation. This model contains two important parameters: $\gamma$ and $\delta$. We, thus, simply call our model $\gamma\delta$CDM. It is distinguished in two important aspects from the $\La$CDM model: 
firstly, the contribution of different energy densities to the Hubble parameter are weighted with different weights, and then, dependence of energy densities to redshift is modified as well. This unorthodox relation of energy content to Hubble parameter brings forth a new way of interpreting the cosmological history. This solution does not allow the existence of a cosmological constant component, however, a dark energy contribution with dependence on redshift is possible. We tested observational relevance of the new solution by best fitting to different data sets. We found that our model could accommodate the idea of cosmological coupling of black holes.
\end{abstract}

%\keywords{Hubble parameter, anisotropy, modified gravity}

\maketitle

%%%%%%%%%%.      Introduction

\section{Introduction}

Advances in cosmology in both observational and theoretical fronts in the last decades were immense and these theoretical and the observational advancements supported each other to build the highly successful ``standard model'' of cosmology, so called $\La$CDM model. Under the assumptions of isotropy and homogeneity of matter distribution in the Universe, and the validity of Einstein's theory of gravity at all classical scales, the $\La$CDM model brings theoretical explanation to observations in both the late and the early universe. Scientific explanation, however, is never complete and scientific progress accelerates after inconsistencies in the paradigm itself start to appear and new ideas are begin to be explored in order to resolve those inconsistencies. 

In cosmology, first of all, there is an important inconsistency between numerically fitted value of the Hubble constant to the late and the early universe data. The discrepancy is on the order of $5\si$ \cite{Abdalla:2022yfr,Shah:2021onj,Freedman:2021ahq} and it has been called a ``crisis'' \cite{Verde:2019ivm}. Whereas Hubble constant is calculated from the Cosmic Microwave Background (CMB) data with a theoretical input \cite{Planck:2018vyg}, it is been fitted to the late universe observational data independent of any cosmological model \cite{Riess:2021jrx,Pesce:2020xfe,Freedman:2019jwv,Wong:2019kwg,Birrer:2018vtm,Dominguez:2019jqc,LIGOScientific:2017adf,Moresco:2017hwt}. Since these discrepancy is not resolved in $\La$CDM model yet \cite{Abdalla:2022yfr,Shah:2021onj,DiValentino:2021izs,Perivolaropoulos:2021jda,Schoneberg:2021qvd,Knox:2019rjx}, it is plausible to seek alternative theoretical explanations to the observational data. One avenue of research has been to change the third assumption of the standard cosmology, namely that the theory of gravitation could be different than the Einstein's theory on scales larger than at least the scale of the Solar System. 

One of the simplest modifications of the Einstein's theory of gravity is the $f(R)$ theory of gravity (see reviews \cite{Sotiriou:2008rp,DeFelice:2010aj,Nojiri:2010wj,Nojiri:2017ncd} and references therein). Action of this theory includes an arbitrary function of the scalar curvature R and in that aspect it is different than the Einstein-Hilbert action, which depends on the scalar curvature itself. There are many works on the cosmological implications of $f(R)$ gravity since Starobinsky's seminal paper \cite{Starobinsky:1979ty,Starobinsky:1980te,Vilenkin:1985md}. In the present work we are going to present a new solution to the field equations of $f(R)$ gravity in the anisotropic Bianchi type I background geometry \cite{Collins:1973lda}. The fact that we work in an anisotropic background is motivated by the ``anomalies" \cite{Buchert:2015wwr,Schwarz:2015cma} in the observational data both in the early and the late universe. An anisotropic universe under the influence of Einstein gravity with a positive cosmological constant tends to isotropize asymptotically \cite{Starobinsky:1962,Wald:1983ky}. In the case of $f(R)$ gravity, however, it is possible to have anisotropically expanding cosmological solutions \cite{Barrow:2005qv}. This is the reason we aim to find a theoretical explanation to the anomalies and inconsistencies in the current cosmological model with an anisotropic background solution in the $f(R)$ theory of gravity.

CMB data explicitly includes anomalies that hint the existence of a global anisotropy. These were persistently observed by the COBE \cite{COBE:1992syq,Kogut:1996us}, WMAP \cite{Bennett:2010jb,deOliveira-Costa:2003utu,WMAP:2003ivt,WMAP:2003zzr,WMAP:2003elm} and Planck experiments \cite{Planck:2015igc,Planck:2019kim}. Among these anomalies the most well known is the lack of power in the quadrupole moment in the CMB power spectrum \cite{Planck:2018vyg,Buchert:2015wwr,Schwarz:2015cma,Campanelli:2006vb,Cea:2022mtf,Campanelli:2007qn,Cea:2019gnu}. Other than that the quadrupole and octupole moments are observed to be aligned with each other and the motion of the solar system, even though they are expected to be independent of each other \cite{Copi:2013jna,Land:2005ad,Schwarz:2004gk}. Additionally there is an observed power asymmetry between northern and southern hemispheres of our chosen celestial sphere \cite{Planck:2019evm,Mukherjee:2015mma,Javanmardi:2016whx,Axelsson:2013mva,Eriksen:2007pc}, and there exists so called cold spots on the microwave sky \cite{Vielva:2003et,Cruz:2006sv}. Each of these anomalies in data might have different reasons and might require different physical solutions. However they might also be collectively pointing out the existence of a preferred direction in space \cite{Schwarz:2015cma,Cea:2022mtf,Luongo:2021nqh,Krishnan:2021jmh,Rodrigues:2007ny,Bridges:2007ne,Migkas:2020fza}. There are further hints of anisotropy in the Pantheon supernova data from the late Universe \cite{Pan-STARRS1:2017jku,Mohayaee:2021jzi,Zhao:2019azy,Amirhashchi:2018nxl,Colin:2019opb,Colin:2010ds}. Such an anisotropic background geometry might be captured by Bianchi type metrics, simplest of which is the Bianchi type I metric \cite{Campanelli:2006vb,Cea:2022mtf,Campanelli:2007qn,Cea:2019gnu,Akarsu:2019pwn,Akarsu:2020pka,Akarsu:2021max,Tedesco:2018dbn,Amirhashchi:2018bic,Hossienkhani:2014zoa,Nojiri:2022idp}. We will not treat $f(R)$ gravity field equations in the anisotropic Bianchi type I background geometry as effective Einstein equations \cite{Capozziello:2005mj,Nojiri:2006ri,Capozziello:2006dj,Faraoni:2018qdr,Sultana:2022qzn,Benisty:2023qcv}. Thus the solution will relate the Hubble parameter to the energy content of the universe very differently compared to the standard general relativistic formalism. Our solution contains two important parameters: $\gamma$ and $\delta$. We, thus, simply call our model $\gamma\delta$CDM.

One of the important aspects of the new solution is that the contribution of different energy densities to the Hubble parameter are weighted with different weights. Other than that, dependence of energy densities to redshift is also modified. This unorthodox relation of the energy content to the Hubble parameter brings forth a new way of interpreting the cosmological history. Our solution does not allow existence of a cosmological constant component, however a dark energy contribution with dynamical dependence on redshift is possible. 
We test the observational relevance of the $\gamma\delta$CDM model by best fitting to different data sets, such as the Pantheon type Ia supernovae data \cite{Pan-STARRS1:2017jku}, the cosmic chronometers (CC) Hubble data \cite{Jimenez:2001gg,Favale:2023lnp} and the Baryon Acoustic Oscillations (BAO) data \cite{Staicova:2021ntm} of the late universe, and the CMB data \cite{Planck:2018vyg} of the early universe.

The important difference between the $\gamma\delta$CDM model and the $\La$CDM model manifests itself in the best fit value of the dark energy density, $\Omega_{e0}$. From the data fit analysis with the above data sets we find that the best fit value for the dark energy density, $\Omega_{e0}$, is just a few percent of the universe's total energy. 
We note that such a \emph{low} value for the density of dark energy could be explained with the cosmological coupling of black holes \cite{Croker:2021duf,Farrah:2023opk,Croker:2020plg}. We find that the best fit value of $(3-k)$, where $k$ is called cosmological coupling constant, inferred by comparing supermassive black holes in five samples of elliptical galaxies at $z > 0.7$ to those in contemporary elliptical galaxies \cite{Farrah:2023opk} is well within the 1$\sigma$ regions of the $\gamma$ parameter for all combinations of the data sets. 
This whole line of argument has a theoretical basis put forward long time ago by Gliner \cite{Gliner:1966}, who argues that the state of collapsed matter in a black hole interior can be considered as a source of dark energy with negative pressure and $p+\rho \geqslant 0$.

This paper is organized as follows: in the next section we will present the action and the field equations of the $f(R)$ gravity. Afterwards, we will write the field equations in terms of the Hubble and the shear anisotropy parameters in the Bianchi type I background. Then, under the assumption that the matter content can be approximated by a perfect fluid, the shear anisotropy parameter will be shown to obey a Gauss--Codazzi type equation. After obtaining redshift dependence of shear anisotropy parameter, subsequent solution of the field equations will provide us a new solution of the Hubble parameter in terms of the energy content of the universe. In section \ref{method} we are going to present the data sets with which we will test our solution and summarize the data fitting method that is used. In section \ref{results} we will present and discuss the relevance of the results of the data fit analysis. Lastly in section \ref{conc} we will summarize our work and comment on our results.

%%%%%%%%%%.       f(R) gravity

\section{Theory \label{theory}}

\subsection{f(R) theory of gravity \label{fR}}

The action of $f(R)$ gravity depends on an arbitrary function of the scalar curvature $R$. In such a theory total action can be written as
\be \label{action}
S=-\int d^4 x \sqrt{-g}\left( \fr1{2\kappa} f(R) + \mathcal{L}_m \right)
\ee
where $\mathcal{L}_m$ denotes the Lagrangian density for matter fields and $\kappa = 8\pi G$ is the Einstein's constant in the natural units (c=1). 

The field equations of $f(R)$ gravity can be obtained by varying the action with respect to the metric. They are given by
\be
f_R R_{\mu\nu} -\fr12 f g_{\mu\nu} +\left( g_{\mu\nu}\Box - \nabla_\mu \nabla_\nu \right) f_R = \kappa T_{\mu\nu}
\label{field}
\ee
where $T_{\mu\nu} = -\fr{2}{\sqrt{-g}} \fr{\delta \mathcal{S}_m}{\delta g_{\mu\nu}}$ is the energy-momentum tensor, and $f_R = \fr{\partial f}{\partial R}$.

%%%%%%%%%%.       Bianchi type I

\subsection{Ellipsoidal Universe}

On the hypothesis that  the quadruple anomaly in CMB data could be due to existence of a preferred direction in space, we adopt an ellipsoidal model of the Universe as in \cite{Campanelli:2006vb,Campanelli:2007qn,Cea:2019gnu}. This model is described by a Bianchi type I metric, which is simply given by
\be \label{B1}
ds^2=-dt^2+A(t)^2dx^2+B(t)^2 (dy^2+dz^2) \, ,
\ee
where directional scale parameters $A(t)$ and $B(t)$ are different from each other. Directional Hubble parameters are defined as $H_x = \dot{A}/A$ and $H_y = H_z = \dot{B}/B$. Physical volume is given by $V(t) = A\cdot B^2$ and we define an average scale parameter $a(t)$ by the relation $V(t) = a(t)^3$.

We then define two new parameters in terms of the directional Hubble parameters as
\be
H (t) = \fr13 ( \frac{\dot{A}}{A} +2\frac{\dot{B}}{B} ) \quad \mathrm{and} \quad S (t) =\frac{\dot{A}}{A} - \frac{\dot{B}}{B} \, .
\ee
Here $H(t)$ is the average Hubble parameter and we call $S(t)$ the shear anisotropy parameter. 
$S^2(t)$ is related to so called shear scalar $\si^2(t)$ that quantifies the anisotropic expansion \cite{Leach2006} by $S^2(t) = 3\si^2(t)$.
Scalar curvature of Bianchi type I spacetime (\ref{B1}) is given by 
\be \label{R}
R = 12H^2 +6\dot{H} +\fr23 S^2 \, .
\ee
We further express the Hubble parameter in terms of the average scale parameter $a(t)$ as $H(t) = \dot{a}/a$, which is the usual relation.

Instead of expressing the field equations in terms of the metric functions $A(t)$ and $B(t)$ we are going to utilize the Hubble and the anisotropy parameters. In terms of $H(t)$ and $S(t)$ the field equations are
\beqa
\mathrm{(00)} &:&  f_R \left( 3 H^2+3 \dot{H}+\frac{2 S^2}{3} \right) -\fr12 f - 3f_{RR} H \dot{R} = -\kappa\rho \, , \label{00} \\
\mathrm{(11)} &:&  f_R \left( 3 H^2+2 H S+\dot{H}+\frac{2 \dot{S}}{3} \right) -\fr12 f 
- f_{RR} \left[ ( 2H - \fr23 S ) \dot{R} + \ddot{R} \right] - f_{RRR} \dot{R}^2 = \kappa p \, , \label{11} \\
\mathrm{(22)} &:&  f_R \left( 3 H^2-H S+\dot{H}-\frac{\dot{S}}{3} \right) -\fr12 f 
- f_{RR} \left[ ( 2H +\fr13 S ) \dot{R} + \ddot{R} \right] - f_{RRR} \dot{R}^2 = \kappa p \, , \label{22}
\eeqa 
where we approximated the matter content as a perfect fluid. This means that there are no anisotropic pressure components and therefore the left hand sides of $(11)$ and $(22)$ field equation components should be the same. In these equations we have $f_{RR} = \fr{\partial^2 f}{\partial R^2}$ and $f_{RRR} = \fr{\partial^3 f}{\partial R^3}$. $(33)$ component of the field equations is the same as the $(22)$ component, thus it is not written.

%Difference of the left hand sides of equations (\ref{11}) and (\ref{22}) should be equated to zero:
Combining the equations (\ref{11}) and (\ref{22}) we obtain
\be \label{ani}
%0 = f_R \left( 2 H S +\frac{2}{3} \dot{S} \right) +\fr23 f_{RR} S\dot{R} 
0 =  f_R S \left( 3 \fr{\dot{a}}a + \fr{\dot{S}}S \right) + f_{RR} S\dot{R} \, ,
\ee
which is equivalent to the trace free Gauss--Codazzi equation \cite{Leach2006,Banik2016} given by
\be \label{GC}
\fr{\dot{S}}S = - \left[ 3 \fr{\dot{a}}a + \fr{\dot{f}_R}{f_R} \right]
\ee

Assuming that the shear anisotropy parameter has the time dependence such as
\be \label{vareps}
\dot{S} = -\varepsilon HS
\ee
one can easily solve equation (\ref{GC}) and determine $f_R = \fr{\partial f}{\partial R}$ and $S(t)$ as a function of the scale parameter $a(t)$ as
\be \label{anis}
f_R  =  \fr\varphi{a^{3-\varepsilon}} \quad and \quad S = \fr{s_0}{a^{\varepsilon}} \, ,
\ee
where $\varphi$ and $s_0$ are the integration constants.

Under the condition (\ref{ani}) the field equations become
\beqa 
\label{iso} \kappa \rho &=& - f_R \left( 3 H^2+3 \dot{H} +\frac{2 S^2}{3} \right) + \fr12 f + 3H \dot{f}_R \, ,  \\
\label{iso1} \kappa p &=&  f_R \left( 3 H^2+\dot{H} \right) - \fr12 f - 2H \dot{f}_R - \ddot{f}_R \, .
\eeqa

%%%%%%%%%%.       Solution

\subsection{Solution to the field equations} 

According to the analysis done in \cite{Leach2006} and \cite{Maartens:1994pb} the contribution of shear to the Hubble parameter would be expected to be different than the case in general relativity. In $f(R)$ theory, the contribution of shear to the Hubble parameter comes due to $2S^2 /3$ term in (\ref{iso}). In general relativity shear dissipates as $S^2 \propto a^{-6}$. In contrast, for example, in the $f(R) = R + \alpha R^2$ theory shear is expected to be dissipated more slowly compared to general relativity \cite{Maartens:1994pb}. Thus we expect parameter $\varepsilon$ of (\ref{vareps}) to be less than three so that $S^2$ dissipates slower than its general relativistic analogue. So we choose that
\be \label{gamma}
\varepsilon = 3 - \delta \quad \mathrm{with} \quad  0 < \delta < 1
\ee
This chosen range of $\delta$ is consistent with the dynamical systems analysis and equation (60) of \cite{Leach2006}.

We assume that the Hubble parameter has a polynomial dependence on the scale parameter $a$ as
\be
H^2 = \sum_\la \frac{h_\la}{a^\la} \, ,
\ee
where $\la \in \mathbb{R}$ are to be determined from the field equations (\ref{iso} \& \ref{iso1}).

We can now solve the function $f(R)$ in terms of the average scale parameter $a$ by using the relations (\ref{anis}) and the form of the scalar curvature R (\ref{R}) in terms of $a$. We find $f(a)$ to have the form
\be \label{f}
f(a) = \varphi \sum_\la \frac{\la (12-3\la)}{\la +\delta} \frac{h_\la}{a^{\la+\delta}} + \varphi \fr{4(3-\delta)}{3(6-\delta)}  \fr{s_0^2}{a^{6-\delta}}- 2\La ,
\ee
where $2\La$ is an integration constant. 

We further assume that there are the usual relativistic and non-relativistic perfect fluid components of the energy density and pressure. Non-relativistic component is the dust (m) with vanishing pressure. Relativistic component is composed of a positive pressure radiation (r) component and a negative pressure dark energy (e) component. We find that the field equations (\ref{iso} \& \ref{iso1}) do not allow a cosmological constant to be a part of $f(R)$ theory, but dark energy component could have equation of state parameter $\omega = \gamma/3 -1$ with $0< \gamma \le 2$. Thus the total energy density and pressure of perfect fluids are given by
\be \label{emt}
\rho = \fr{\rho_{e0}}{a^\gamma} + \fr{\rho_{m0}}{a^3} + \fr{\rho_{r0}}{a^4} \quad \mathrm{and} \quad p = (\fr\gamma3 -1) \fr{\rho_{e0}}{a^\gamma} + \fr13 \fr{\rho_{r0}}{a^4} \, ,
\ee
where $\rho_{e0}, \rho_{m0}$ and $\rho_{r0}$ are the present day dark energy, non-relativistic matter (dust) and relativistic matter (radiation) densities, respectively.

Substituting relations (\ref{anis}, \ref{f} \& \ref{emt}) into the field equations we are able to determine possible $\la \in \mathbb{R}$ values and solve the Hubble parameter in terms of the perfect fluid densities as
\be
3H^2 = \fr{\kappa \rho_e}{\varphi b_\gamma} \fr1{a^{\gamma - \delta}} + \fr{\kappa \rho_m}{\varphi b_3} \fr1{a^{3 - \delta}} + 
\fr{\kappa \rho_r}{\varphi b_4} \fr1{a^{4 - \delta}} + \fr13 \fr{s_0^2}{1-\delta} \fr1{a^{6 - 2\delta}} \, ,
\ee
where coefficients $b_n$ ($n = \gamma,3,4$) are given by
\be \label{bn}
b_n = - \left( 1+ \delta - \fr1{2n} (n -\delta)(4+\delta) \right) \, .
\ee
Coefficient $b_0$ is divergent unless $\delta = 0$, which is the case of general relativity. Therefore we cannot include a cosmological constant term into this model, and thus we take integration constant in the relation (\ref{f}) to be vanishing.

To compare theoretical form of the Hubble parameter with the observational data we express $H^2$ in terms of dimensionless density parameters as
\be \label{H2}
H^2 = \fr{H_0^2}{\varphi} \left[ \fr{\Omega_{e0}}{b_\gamma} \fr1{a^{\gamma - \delta}} + \fr{\Omega_{m0}}{b_3} \fr1{a^{3 - \delta}} + 
\fr{\Omega_{r0}}{b_4} \fr1{a^{4 - \delta}} + \fr{\Omega_{s0}}{1 -\delta} \fr1{a^{6 - 2\delta}} \right] 
\quad \mathrm{with} \quad \Omega_{s0} = \fr{\varphi s_0^2}{9 H_0^2} \, .
\ee
where $\Omega_{s0}$ gives the present day contribution of anisotropic shear to the Hubble parameter and $H_0$ is the Hubble constant at the present time. Other dimensionless present day density parameters ($\Omega_{e0}$ for dark energy, $\Omega_{m0}$ for dust and $\Omega_{r0}$ for radiation) are defined by dividing respective densities with the critical density $\rho_c = 3H_0^2/8\pi G$.

The general relativistic limit is obtained as $\delta \rightarrow 0$ and $\varphi \rightarrow 1$. In that case we find that $b_\gamma = b_3 = b_4 = 1$ and that $\Omega_{e0} + \Omega_{m0} + \Omega_{r0} + \Omega_{s0} = 1$ as it should be for a flat universe described by general relativity and Friedmann equations. We also note that in the same limit shear dissipates with the sixth power of the average scale parameter ($a^{-6}$) as is expected in the general relativistic case.

For the sake of simplicity we will not have $\varphi$ as a free parameter and take $\varphi = 1$ also in the general $f(R)$ case when we compare our solution with the cosmological data. A non-vanishing $\delta$ parameter makes all the important difference between these theory and the general relativity. Value of $\delta$ parameter determines both the contribution of different perfect fluid components to the expansion of the Universe and also how these components are dissipates as the Universe expands. Coefficients $b_n$ for $n = \gamma,3,4$ depend on the value of $\delta$ and they weight the contributions of different $\Omega_i$ to the Hubble parameter. Depending on the value of $\delta$ the contribution of one component may get enhanced as the contribution of another component is diminished. 

Another important effect of non-zero value of $\delta$ is that even though perfect fluid components diminish with expansion depending on their physical nature, their contributions to the expansion diminishes differently. For example dust (non-relativistic matter) component diminishes with $a^3$, but its contribution to the expansion diminishes slower with $a^{3-\delta}$. The same also is true for the other perfect fluid components. Their contributions to the Hubble parameter diminishes slower than their actual physical dynamics requires.

Faster dependence on the scale parameter was observed in solutions to the field equations of Brans-Dicke theory with non-zero Brans-Dicke parameter $\omega_{BD}$ in the case of a flat isotropic universe in \cite{Boisseau:2010pd} and a flat anisotropic universe in \cite{Akarsu:2019pvi}. As is well known, Brans-Dicke theory with a kinetic term for the auxiliary scalar field is not equivalent to any f(R) theory \cite{Velasquez:2018euw}. Thus, the theories of gravity studied in those works are completely different from the theory we study here. Both of these works have exact solutions if the perfect fluid is only composed of non-relativistic pressureless fluid, i.e., dust, and also both of them cannot include the radiation component in the exact analytical solution and can only include ``effective'' dark energy. This is also the case in \cite{Schiavone:2022wvq}. Thus, these models \cite{Boisseau:2010pd,Akarsu:2019pvi,Schiavone:2022wvq} are some unrealistic solutions in the scalar-tensor theory framework. Additionally, in these works, the dust component's contribution to the Hubble parameter diminishes faster than its actual physical dynamics requires. This is in complete contrast with what we have here in the f(R) theory framework. 

Our solution (\ref{H2}) has an enhanced ability to adjust itself to fit with different data sets from completely different cosmological eras. It is obvious that existence of the parameters $\delta$ and $\gamma$ cause the essential difference of this model with the $\La$CDM or related anisotropic $\La$CDM models \cite{Akarsu:2019pwn,Akarsu:2021max}. Therefore we call this model simply as $\ga\ld$CDM model. In the rest of this paper we are going to summarize the data analysis and check the expectations of the $\ga\ld$CDM model (\ref{H2}) with various cosmological data sets.

%%%%%%%%%%.       Data

\section{Data and Methodology \label{method}} 

\subsection{Data}

To constrain our parameters in the theoretical model with the observations we use the latest and most relevant data sets. The following four data sets are distinct both in the physical origin and the observational methodology:

\ben

\item CC Hubble data: The latest compilation of data obtained through the cosmic chronometers method \cite{Jimenez:2001gg} comprise 32 data points which are given in Table \ref{HCC} (see references therein) with low redshift ($0.07 < z < 1.97$). These redshift values are the same in \cite{Favale:2023lnp}; however, the values of H(z) are taken from those calculated by the BC03 model for the redshifts, which are mentioned in \cite{Moresco:2012,Moresco:2015,Moresco:2016}.

For the data points taken from \cite{Moresco:2012,Moresco:2015,Moresco:2016} there are contributions to the model covariance matrix due to the uncertainties due to star formation history, the IMF, the stellar library, and the stellar population synthesis model. Detailed information and the way we follow the calculation of the covariance matrix is the same as in \cite{Moresco:2020}. The rest of the data points in Table \ref{HCC} are uncorrelated with the data taken from \cite{Moresco:2012,Moresco:2015,Moresco:2016}, and, therefore, we include them diagonally in the covariance matrix. 

The chi-squared function for the 32 H(z) measurements, denoted by $\chi^2_{CC}$, is defined as
\be
\chi^2_{CC} = M^T Cov^{-1}M,
\ee
where M represents the residual between model prediction $H^{th}(z_i)$ and  observational data $H^{obs}(z_i)$ as
\be
M=H^{obs}(z_i)-H^{th}(z_i)\ .
\ee

\begin{table}[hbt!]
    \centering
    \begin{tabular}{lcl}
    \hline
     $z_i$ & $H^{obs}(z_i)\pm\sigma(z_i)$ & \qquad References \\
     \hline
     \hline
     0.07 & 69 $\pm$ 19.6 & \cite{Zhang:2014} Zhang et al. (2014) \\
     \hline
     0.09 & 69 $\pm$ 12 & \cite{Jimenez:2003} Jimenez et al. (2003) \\
     \hline
     0.12 & 68.6 $\pm$ 26.2 & \cite{Zhang:2014} Zhang et al. (2014) \\
     \hline
     0.17 & 83 $\pm$ 8 & \cite{Simon:2005} Simon et al. (2005)\\
     \hline
     0.1791 & 74.91 $\pm$ 3.8 & \cite{Moresco:2012} Moresco et al. (2012)\\
     \hline
     0.1993 & 74.96 $\pm$ 4.9 & \cite{Moresco:2012} Moresco et al. (2012) \\
     \hline
     0.2 & 72.9 $\pm$ 29.6 & \cite{Zhang:2014} Zhang et al. (2014) \\
     \hline
     0.27 & 77 $\pm$ 14 &  \cite{Simon:2005} Simon et al. (2005) \\
     \hline
     0.28 & 88.8 $\pm$ 36.6 & \cite{Zhang:2014} Zhang et al. (2014) \\
     \hline
     0.3519 & 82.78 $\pm$ 13.94 & \cite{Moresco:2012} Moresco et al. (2012) \\
     \hline
     0.3802 & 83 $\pm$ 13.54 & \cite{Moresco:2016} Moresco et al. (2016) \\
     \hline
     0.4 & 95 $\pm$ 17 & \cite{Simon:2005} Simon et al. (2005) \\
     \hline
     0.4004 & 76.97 $\pm$ 10.18 & \cite{Moresco:2016} Moresco et al. (2016) \\
     \hline
     0.4247 & 87.08 $\pm$ 11.24 & \cite{Moresco:2016} Moresco et al. (2016) \\
     \hline
     0.4497 & 92.78 $\pm$ 12.9 & \cite{Moresco:2016} Moresco et al. (2016) \\
     \hline
     0.47 & 89 $\pm$ 49.6 & \cite{Ratsimbazafy:2017} Ratsimbazafy et al. (2017) \\
     \hline
     0.4783 & 80.91 $\pm$ 9.044 & \cite{Moresco:2016} Moresco et al. (2016) \\
     \hline
     0.48 & 97 $\pm$ 62 & \cite{Stern:2010} Stern et al. (2010) \\
     \hline
     0.5929 & 103.8 $\pm$ 12.49 & \cite{Moresco:2012} Moresco et al. (2012) \\
     \hline
     0.6797 & 91.6 $\pm$ 7.96 & \cite{Moresco:2012} Moresco et al. (2012) \\
     \hline
     0.75 & 98.8 $\pm$ 33.6 & \cite{Borghi:2022} Borghi et al. (2022) \\
     \hline
     0.7812 & 104.5 $\pm$ 12.19 & \cite{Moresco:2012} Moresco et al. (2012) \\
     \hline
     0.8754 & 125.1 $\pm$ 16.7 & \cite{Moresco:2012} Moresco et al. (2012) \\
     \hline
     0.88 & 90 $\pm$ 40 & \cite{Stern:2010} Stern et al. (2010) \\
     \hline
     0.9 & 117 $\pm$ 23 & \cite{Simon:2005} Simon et al. (2005) \\
     \hline
     1.037 & 153.7 $\pm$ 19.67 & \cite{Moresco:2012} Moresco et al. (2012) \\
     \hline
     1.3 & 168 $\pm $17 & \cite{Simon:2005} Simon et al. (2005) \\
     \hline
     1.363 & 160 $\pm$ 33.58 & \cite{Moresco:2015} Moresco (2015)  \\
     \hline
     1.43 & 177 $\pm$ 18 & \cite{Simon:2005} Simon et al. (2005) \\
     \hline
     1.53 & 140 $\pm$ 14 & \cite{Simon:2005} Simon et al. (2005) \\
     \hline
     1.75 & 202 $\pm$ 40 & \cite{Simon:2005} Simon et al. (2005) \\
     \hline
     1.965 & 186.5 $\pm$ 50.43 & \cite{Moresco:2015} Moresco (2015)  \\
     \hline
     \hline
\end{tabular}
    \caption{Cosmic Chronometers Hubble Data. $H^{obs}(z_i)\pm\sigma(z_i)$ is in $km/s/Mpc$.}
    \label{HCC}
\end{table}

\item CMB data: Following \cite{Amirhashchi:2019jpf} we only use CMB distance measurements in the Bayesian analysis.
The first peak of the CMB spectrum, denoted by $l_\ast$, is called the angular scale of the sound horizon at the last scattering surface and is given by
\be \label{ls}
l_\ast=\pi\frac{D_M(z_\ast)}{r_s(z_\ast)}\ ,
\ee
where $z_\ast=1089.9$ \cite{Planck:2018vyg} is the redshift at the last scattering surface.
The comoving angular diameter distance to the last scattering surface $D_M(z_\ast)$ and the comoving sound horizon $r_s(z_\ast)$ are given by
\begin{equation} \label{r_s}
D_M (z_\ast) = \int_{0}^{z_{\star}} \frac{dz}{H(z)}\quad \mathrm{and} \quad 
r_s (z_\ast) = \int\limits_{z_\ast}^{\infty} \frac{c_s (z)}{H(z)} dz \ ,
\end{equation}
where $c_s=\frac{c}{\sqrt{3(1+R_b/(1+z))}}$ is the sound speed in the photon-baryon fluid and $R_b/(1+z)=3\rho_{b0}/(4\rho_{r0})$ is the ratio of baryon to photon momentum density \cite{Eisenstein:1997, Wang:Yun}. Given the observed average temperature of the CMB, $T_{CMB}=2.7255\ K$, $R_b$ is given by $R_b=31500\Omega_bh^2(T_{CMB}/2.7)^{-4}$ with $\Omega_b h^2=0.02226$ \cite{Planck:2015}.

Another parameter related to CMB power spectrum is the shift parameter R, which is related with the overall CMB acoustic peak amplitude \cite{Amirhashchi:2019jpf}:
\begin{equation} \label {R}
R=H_0\sqrt{\Omega_m}\int\limits_{0}^{z_\ast} \frac{dz}{H(z)}
\end{equation}
The mean values of angular scale and shift parameters are $l_\ast=301.76$ and $R=1.7488$ with standard deviations $\sigma_l = 0.14$ and $\sigma_R = 0.0074$ \cite{Planck:2015,Amirhashchi:2019jpf}. 

The chi-squared function of the CMB distance data is defined in terms of a covariance matrix as
\begin{equation}
\chi^2_{CMB}= D^T C^{-1} D\ , \quad \mathrm{with} \quad C^{-1} = \begin{pmatrix}
1.412 & -0.762 \\
-0.762 & 1.412 \\
\end{pmatrix}\ ,
\end{equation}
where D is a column matrix containing residuals between model predictions (\ref{ls},\ref{R}) and observational values.

\item BAO data: 
The baryon acoustic oscillations are density fluctuations of the visible baryonic matter.
We use BAO measurements by surveys such as SDSS, DES and WiggleZ as summarized in Table 1 of \cite{Staicova:2021ntm} comprising 18 data points with low redshift $(0.11 \leq z \leq 2.4)$. 
The BAO measurements constrain the ratio $d_b=D_A/r_d$ of the angular diameter distance $D_A$ to the baryon drag epoch and the comoving size of the sound horizon at the baryon drag $r_d = r_s(z_d)$. These quantities are given by
\begin{equation} \label{r_s}
D_A (z_d)= \frac{c}{(1+z)} \int\limits_0^{z_d} \frac{dz}{H(z)}\quad \mathrm{and} \quad 
r_s (z_d) = \int\limits_{z_d}^{\infty} \frac{c_s (z)}{H(z)} dz \ .
\end{equation}

Then the chi-squared function for the BAO measurements is given as
\begin{equation}
    \chi^2_{BAO}=\left( \frac{d_b^{obs}-d_b^{th}}{\sigma_b} \right)^2
\end{equation}
in terms of observational and theoretical differences of the ratio $d_b$.
\newline

\item Pantheon+ data: 
We also use the Type Ia SNe distance modulus measurements from the Pantheon+ sample \cite{Brout:2022vxf,Scolnic:2021amr} to constrain our cosmological model parameters. The Pantheon+ sample ranges in redshift $0.001 < z < 2.26$ and includes 1701 light curves of 1550 distinct Type Ia supernovae. All these data can be found in the GitHub repositories of \cite{Scolnic:pant}. We exclude data with redshift $z<0.01$ for the peculiar velocities of supernovae with low redshifts not to affect our analysis.
Thus, we use the data in the redshift range $0.01 < z < 2.26$. The Pantheon+ sample also incorporates SH0ES \cite{Riess:2021jrx} Cepheid host distances to constrain the absolute magnitude parameter $M$. 

The distance residuals are split into two groups \cite{Brout:2022vxf}:
\begin{equation} \label{res}
\Delta \mu_i =
\begin{cases} 
    \mu_{\text{dat},i} - \mu_i^{\text{Cepheid}} & \text{, if } i \in \text{Cepheid hosts} \\
    \mu_{\text{dat},i} - \mu_{\text{model}}(z_i) & \text{, otherwise}\ .
\end{cases}
\end{equation}
The observed distance modulus is given by
\begin{equation}
\mu_{dat}=m_b ^{corr}-M.
\end{equation}
The standardized and adjusted $m_b^{corr}$ magnitudes are available in the \texttt{`m\_b\_corr'} column of the Pantheon+ dataset \cite{Scolnic:pant}. The distance modulus, which depends on the model, is defined as \cite{Lovick:2023tnv}
\begin{equation}
\mu_{model}=5\log_{10}\left(\frac{d_L}{10\text{pc}}\right)
\end{equation}
where $d_L$ represents the luminosity distance, incorporating the model parameters. It is given by \cite{Alonso-Lopez:2023hkx}
\begin{equation}
d_L=c \left(1+z_{hel}\right) \int\limits^{z_{\text{HD}}} _0\frac{dz'}{H(z')} \ .
\end{equation}
Here, $z_{\text{hel}}$ denotes the heliocentric redshift of supernovae, while $z_{HD}$ represents the redshift in the CMB frame with corrections for peculiar velocities. These values are extracted from the  \texttt{`zhel'} and \texttt{`zHD'} columns in the Pantheon+ dataset \cite{Scolnic:pant}.

The absolute magnitude, denoted by $M$, is a parameter unaffected by the cosmological model and instead determined by the dynamics of the supernova. There exists a degeneracy between $H_0$ and $M$. To resolve this, we utilize the SH0ES Cepheid distance measurements (identified by `\texttt{IS\_CALIBRATOR=1}' in the data file \cite{Scolnic:pant}). We treat the absolute magnitude $M$ as a variable parameter, with a prior distribution assumed to be uniform within the range [-20, -18] \cite{Lovick:2023tnv,Alonso-Lopez:2023hkx}.
 
Data points with redshifts below 0.01 are excluded, resulting in 1580 remaining data points for the Hubble Diagram and 10 for Cepheid distance measurements (identified by `\texttt{CEPH\_DIST}' in the dataset \cite{Scolnic:pant})).

The chi-squared function for the Pantheon+ data is formulated as
\begin{equation}
\chi^2= \Delta \mu^TC^{-1}\Delta\mu
\end{equation}
where $\Delta \mu$ denotes the vector of distance modulus residuals as specified in equation (\ref{res}), and $C$ represents the covariance matrix incorporating both statistical and systematic covariance matrices \cite{Scolnic:pant}.

\een

%%%%%%%%%%.       Methodology

\subsection{Methodology}

In this work we use the Bayesian inference method for parameter estimation and model comparison \cite{Padilla:2019, Hogg:2010}. Bayes theorem is a consequence of the probability axioms. If $x_1$ and $x_2$ are two mutually exclusive events, the probability of both events occurring is equal to the product of the probability of $x_1$ and the probability of $x_2$ given that $x_1$ has already happened:
\be
P(x_1\cap x_2)=P(x_1)P(x_2 | x_1)
\ee
This is also true other way around and gives us,
\be
P(x_1 | x_2)=\frac{P(x_2 | x_1)P(x_1)}{P(x_2)}.
\ee
The equation above can be converted for a given model $\mathcal{M}$, the data $I$ and the parameters $\theta$ in Bayesian inference as
\be \label{bi}
P(\theta | \mathcal{I},\mathcal{M})=\frac{P(\mathcal{I} | \theta,\mathcal{M}) P(\theta | \mathcal{M})}{P(\mathcal{I} | \mathcal{M})}\ ,
\ee
where $P(\theta | \mathcal{I},\mathcal{M})$ is called the posterior probability of a given model $\mathcal{M}$.  
$P(\mathcal{I} | \theta,M)=\mathcal{L}(\mathcal{I} | \theta,\mathcal{M})$, which is known as the likelihood function of the model, is given by
\be \label{Like}
\mathcal{L}(\mathcal{I} | \theta,\mathcal{M}) \propto \exp{\left[-\frac12 \chi^2 (\mathcal{I} | \theta,\mathcal{M})\right]}
\ee
where $\chi^{2} (\mathcal{I} | \theta,\mathcal{M})$ is the chi-squared function. Thus, the minimum value of the chi-squared function gives the maximum likelihood.

We derive posterior probability distributions and the maximum likelihood function with the nested sampling Monte Carlo algorithm MLFriends \cite{Buchner:2014,Buchner:2017} using the UltraNest\footnote{\url{https://johannesbuchner.github.io/UltraNest/}} package \cite{Buchner:2021}. Nested sampling is a Monte Carlo algorithm that computes an integral over the model parameters. In this work, we tried to constrain model parameters and compare the model to the $\Lambda$CDM model by using the UltraNest.
When we use multiple data sets together we calculate the joint chi-squared function.
Then the likelihood function is multivariate joint Gaussian likelihood given by (\ref{Like}).

The free parameters of our cosmological model are $H_0,\ \Omega_{m0},\ \Omega_{s0},\ \gamma$ and $\delta$ (\ref{H2}). For $\Omega_{r0}$ we used the value $\Omega_{r0}=h^{-2}.2.469.10^{-5}(1+\frac{7}{8}(\frac{4}{11})^{4/3}N_{eff})$ where $N_{eff}=3.046$ as given in \cite{Mukhanov:2005sc} determined by the value of CMB temperature and the Stefan-Boltzmann law. We fixed $\Omega_{e0}$ in terms of other parameters at $a(t_0) = a_0 = 1$ and constrained the remaining five cosmological parameters through the data analysis. 

Priors are chosen so as to scan the parameter space thoroughly and determine the best fit value of the parameters. The uniform prior distributions are chosen for the parameters $H_0$ and $\Omega_{m0}$, given by $0.5 \leqslant H_0 \leqslant 0.85$ and $0.0 \leqslant \Omega_{m0} \leqslant 1.0$, respectively. For $\Omega_{s0}$ we choose logarithmic prior distribution given by $-16 \leqslant \log_{10} \Omega_{s0} \leqslant 0.0$ due to its expected very low value.
For the analysis summarized in Table \ref{tmv} logarithmic prior distribution is chosen for $\gamma$ with $0.001 \leqslant \gamma \leqslant 3.0$ and uniform prior distribution is chosen for $\delta$ with $0.0 \leqslant \delta \leqslant 0.772$. The corresponding likelihood and contour plots are given in Fig. \ref{fmv}. 
All plots are produced by GetDist, which is commonly used to analyze Monte Carlo samples \cite{Lewis:2019xzd}. 

In section \ref{theory} we emphasized that in this model the dark energy component cannot be a cosmological constant and thus the parameter $\gamma$ cannot be zero (\ref{H2}). Due to this fact, the lower bound of the prior distribution for $\gamma$ could not be zero, but could be arbitrarily close to zero. Since lower bound could be arbitrarily small, but not zero, we use logarithmic prior distribution for the parameter $\gamma$. We have checked various choices for the lower bound of the prior distribution for $\gamma$ and we find that the median value and the minimum chi-squared function for those fits do not change significantly. In the next section we report the results for minimum $\gamma$ chosen to be $0.001$. The lower bound of the prior distribution for $\delta$ is chosen zero, since this value corresponds to the case of general relativistic solution. The upper bound is chosen to be $\frac12 (\sqrt{73}-7) \approx 0.772$ because for larger values of $\delta$ the coefficient $b_3$ (\ref{bn}) becomes negative, which will render the contribution of the matter density to the expansion of the universe to be negative, which, needless to say, is unphysical. 

%%%%%%%%%%.       Results and Discussion

\section{Results and Discussion \label{results}} 

\begin{figure*}[hbt!]
\centering
\begin{subfigure}
    \centering 
    \includegraphics[width=8cm]{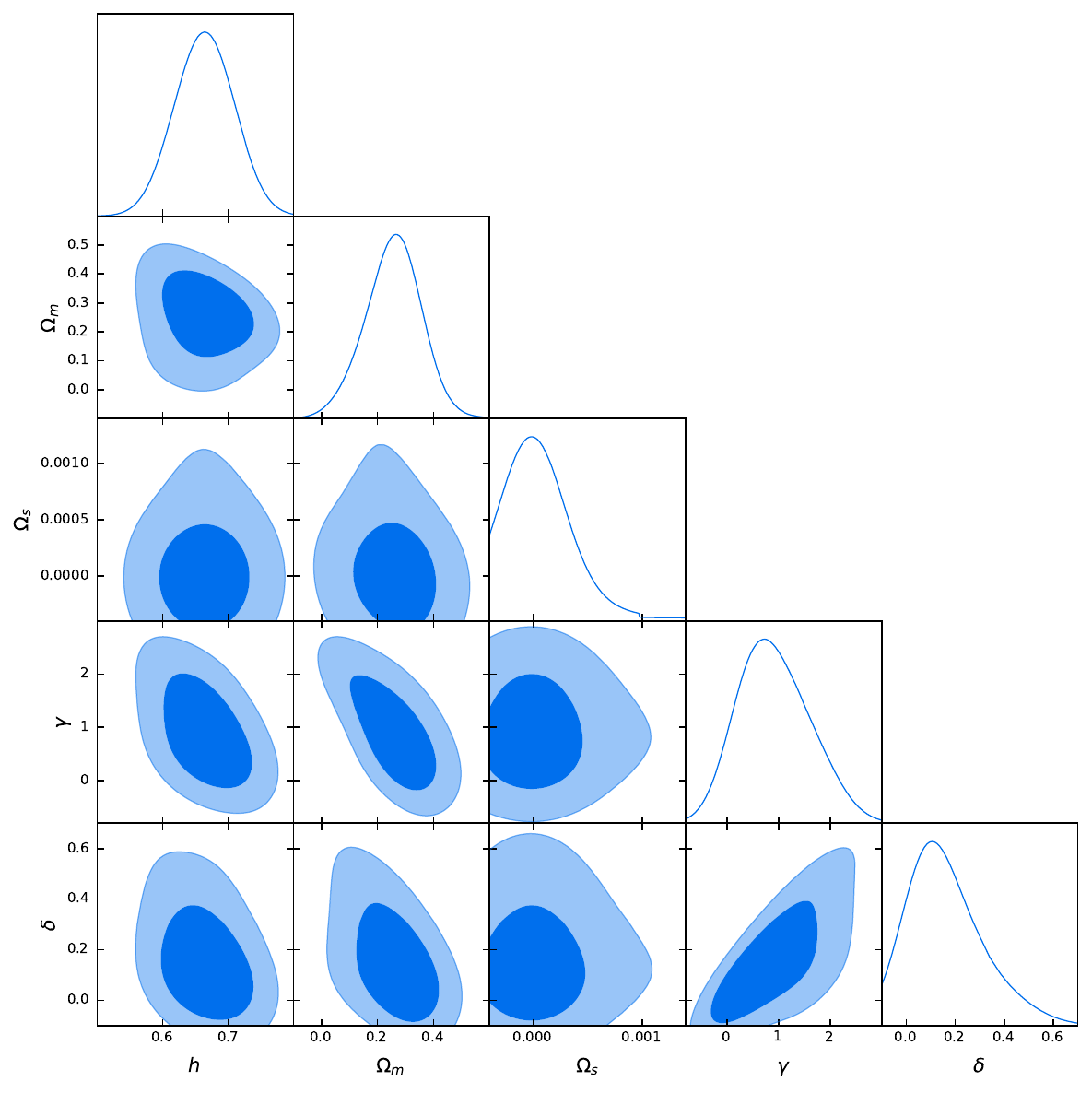}
    \label{fig:sub1}
\end{subfigure}
\begin{subfigure}
    \centering 
    \includegraphics[width=8cm]{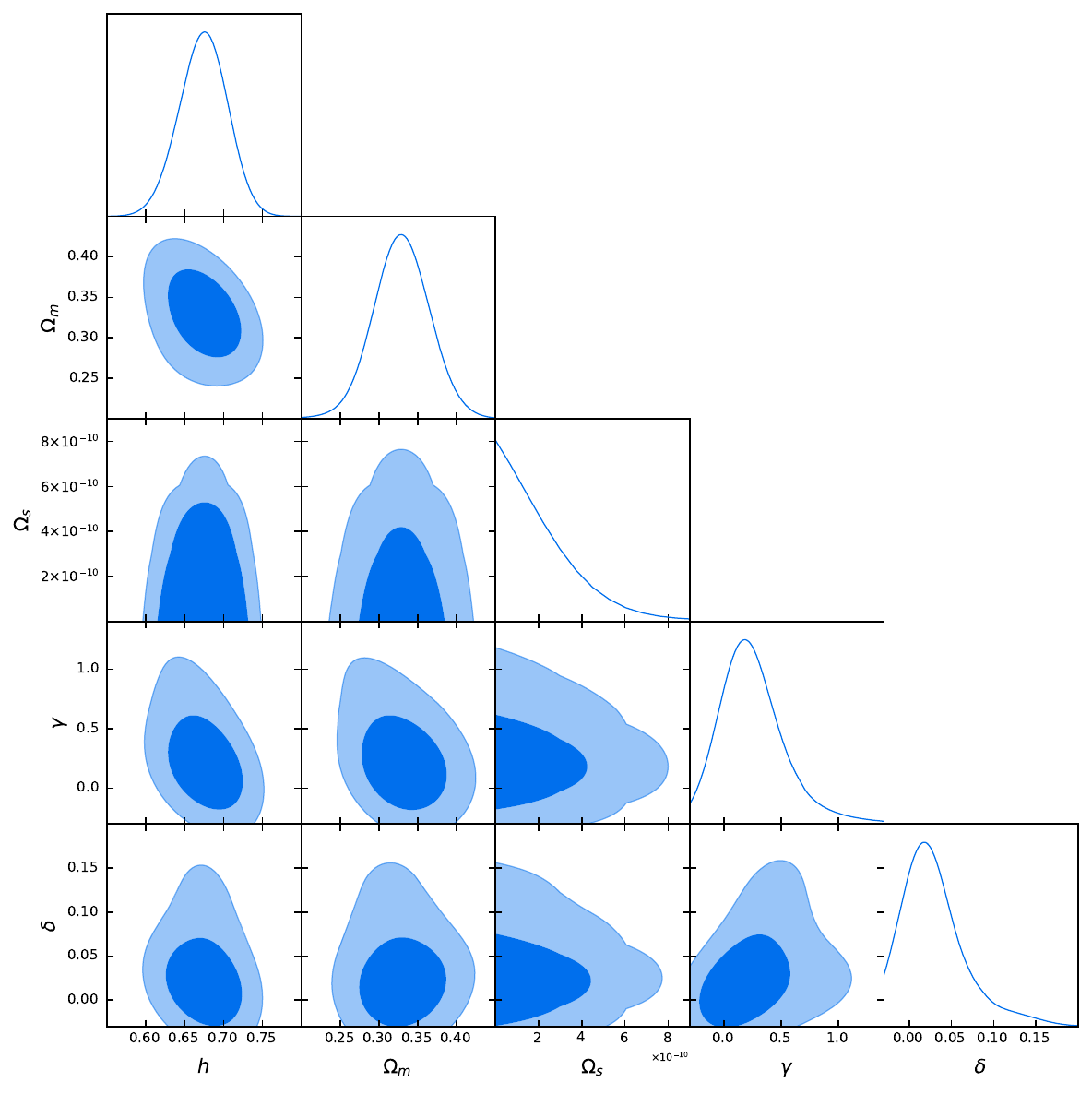}
    \label{fig:sub2}
\end{subfigure}
\begin{subfigure}
    \centering 
    \includegraphics[width=8cm]{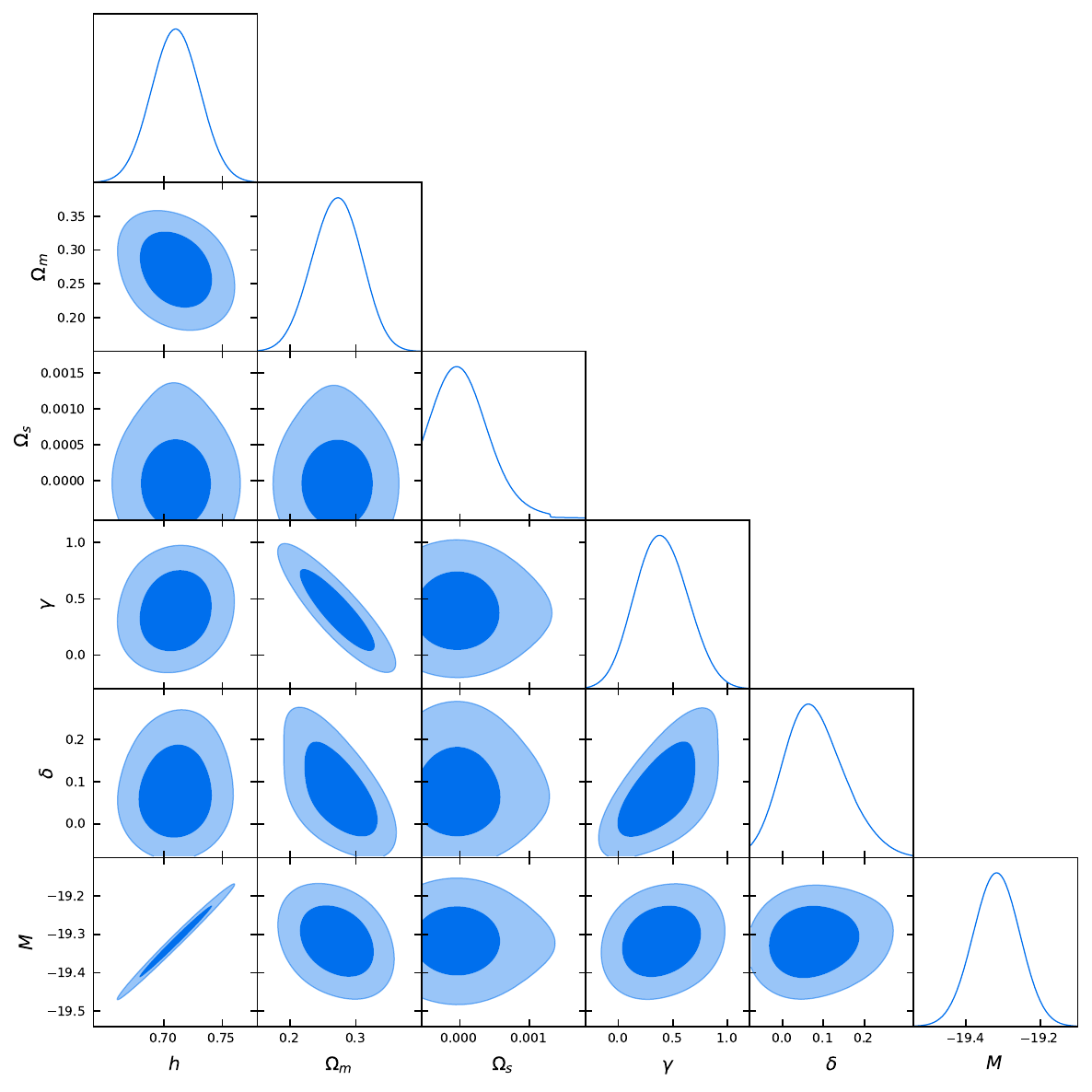}
    \label{fig:sub1}
\end{subfigure}
\begin{subfigure}
    \centering 
    \includegraphics[width=8cm]{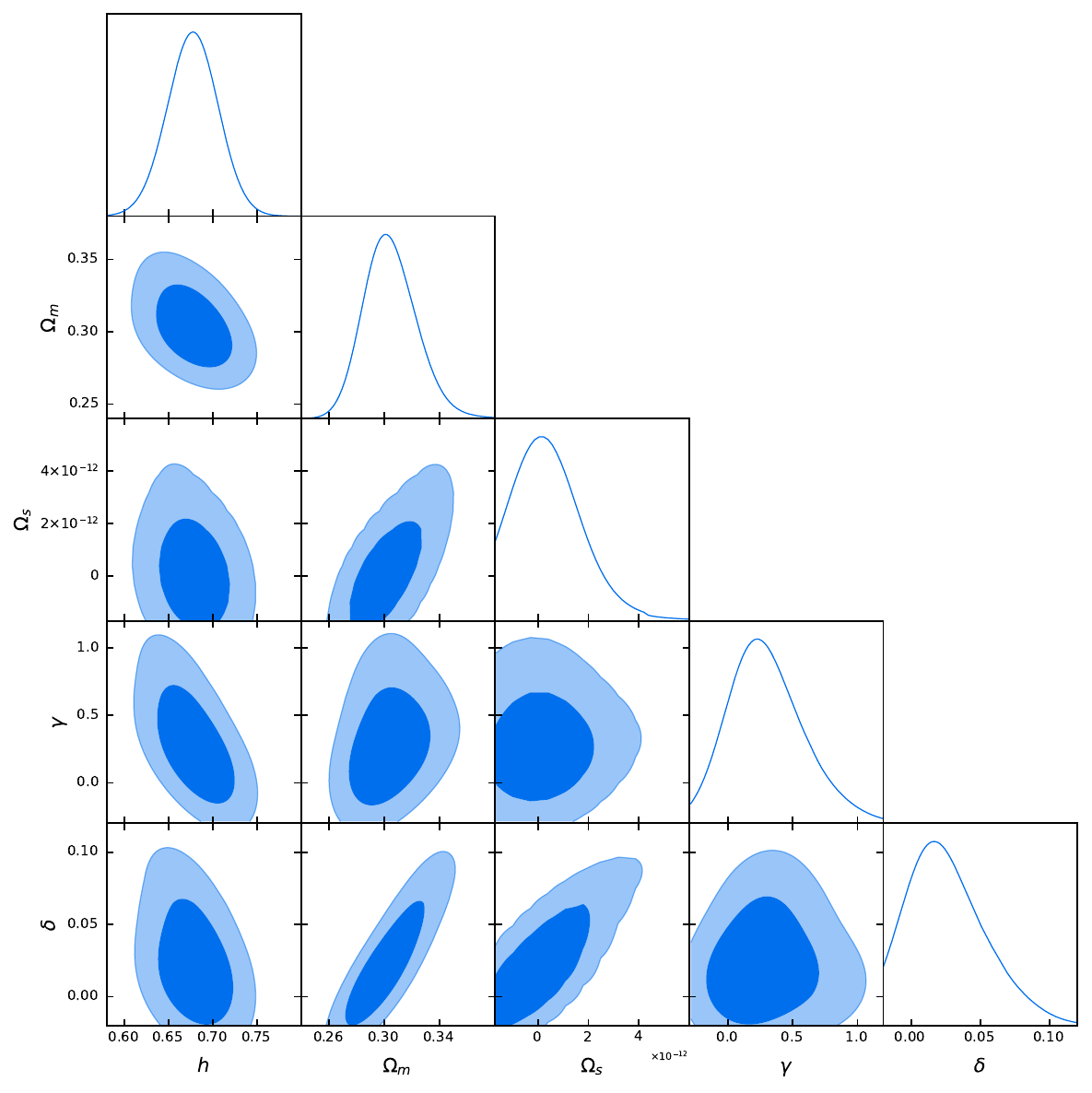}
    \label{fig:sub2}
\end{subfigure}
\caption{The 1D and 2D posterior distributions with 1$\sigma$ and 2$\sigma$ marginalized confidence regions of model parameters from CC (top-left), CC+Pan+ (bottom-left), CC+BAO (top-right), and CC+BAO+CMB (bottom-right) data combinations. The Hubble constant $H_{\rm 0} = 100\cdot h$ is in units of km/s/Mpc.}
\label{fmv}
\end{figure*}

We present the constraints with 68\% CL on the free ($H_0,\ \Omega_{m0},\ \Omega_{s0},\ \gamma$ and $\delta$) and some derived parameters ($\Omega_{e0}, \Omega_{e0} / b_\gamma$ and $\Omega_{m0} / b_3$) of the $\ga\ld$CDM model for different data set combinations in Table \ref{tmv}, and 1D and 2D posterior distributions, with 1$\sigma$ and 2$\sigma$ marginalized confidence regions, for the free parameters in Fig. \ref{fmv}. For comparison we also provide marginalized constraints for the parameters of the $\La$CDM standard model (for which $\delta,\ \gamma$ and $\Omega_{s0}$ all vanish) in Table \ref{tmv}. We also present best fit values of the cosmological parameters in Table \ref{tbd}.

\begin{table*}[hbt!]
\centering
\def\arraystretch{1.3}
\begin{tabular}{|l|c|c|c|c|}
\hline 
\hline 
\textbf{Data set} & CC & CC+Pan+ & CC+BAO & CC+BAO+CMB \\
\hline 
\textbf{Model} & $\ga\ld$CDM & $\ga\ld$CDM & $\ga\ld$CDM & $\ga\ld$CDM \\
 & \textcolor{blue}{$\La$CDM} & \textcolor{blue}{$\La$CDM}
 & \textcolor{blue}{$\La$CDM} & \textcolor{blue}{$\La$CDM} \\
\hline
%\hline
$\mathbf{H_0}$
 & $66.39\pm 4.33$ & $71.02\pm 1.89$ & $67.40\pm 2.84$ & $67.72\pm 2.64$  \\
 & \textcolor{blue}{$68.79\pm 4.05$} & \textcolor{blue}{$70.53\pm 1.89$}
 & \textcolor{blue}{$71.17\pm 1.22$} & \textcolor{blue}{$69.53\pm 0.62$}   \\
%\hline
$\mathbf{\Omega_{m0}}$
 &\; $0.257\pm 0.095$\; &\; $0.271\pm 0.034$\; &\; $0.329\pm 0.034$\; &\; $0.304\pm 0.017$\; \\
 &\; \textcolor{blue}{$0.324\pm 0.063$}\; &\; \textcolor{blue}{$0.326\pm 0.017$}\;
 &\; \textcolor{blue}{$0.309\pm 0.017$}\; &\; \textcolor{blue}{$0.286\pm 0.008$}\;  \\
%\hline
$\log_{10}(\mathbf{\Omega_{s0}})$ & $< -3.98$ & $< -3.92$ & $< -11.24$ & $< -12.28$ \\
%\hline
$\mathbf{\gamma}$
 & $0.922\pm 0.656$ & $0.398\pm 0.215$ & $0.261\pm 0.258$ & $0.308\pm 0.259$ \\
%\hline
$\mathbf{\delta}$
 & $0.161\pm 0.145$ & $0.082\pm 0.066$ & $0.031\pm 0.035$ & $0.027\pm 0.026$ \\
%\hline
\hline
\textbf{min} $\mathbf{\chi}^2$ & $14.50$ & $1406$ & $36.31$ & $36.83$ \\
 & \textcolor{blue}{$14.50$} & \textcolor{blue}{$1407$} & \textcolor{blue}{$37.90$} 
 & \textcolor{blue}{$40.31$}  \\
\hline
\hline
\end{tabular} 
\caption{\rm Posterior constraints, with 68\% CL, on the free ($H_0,\ \Omega_{m0},\ \Omega_{s0},\ \gamma$ and $\delta$) parameters of the $\gamma\delta$CDM model for different data set combinations. For comparison we also provide constraints on the parameters of the $\La$CDM model (for which $\delta,\ \gamma$ and $\Omega_{s0}$ all vanish) in blue. Minimum values of the chi-squared functions for both models are also presented.} 
\label{tmv}
\end{table*}

As we examine Fig. \ref{fmv} we note that the posterior distributions for the $\gamma$ and $\delta$ parameters are not Gaussian. Therefore, the mean and the median values for these parameters are quite different than the best fit values (maximum likelihood) as can be observed in Tables \ref{tmv} and \ref{tbd}. Other free parameters ($H_0$ and $\Omega_{m0}$) do have Gaussian distributions and thus their mean values are very close to the best fit values. When we examine the minimum values for the chi-squared functions, we observe that in the case of maximum likelihood $\ga\ld$CDM model is favored over the $\La$CDM model depending on the data set. We note that the CC Hubble data \cite{Jimenez:2001gg,Favale:2023lnp} is not sufficient to constrain the model well. Having only 32 data points does not provide a statistically significant result as 1$\sigma$ regions of the posterior values are quite large for this data set. CC Hubble data set together with the Pantheon+ data set \cite{Pan-STARRS1:2017jku} constrain the model better and give posterior mean value for the Hubble constant in the 2$\sigma$ region of the SH0ES collaboration's result \cite{Riess:2021jrx}. In contrast the data sets that are related to the early universe, BAO \cite{Staicova:2021ntm} and CMB \cite{Planck:2018vyg}
data sets, together with the CC Hubble data set give posterior mean value for the Hubble constant in the 1$\sigma$ region of the Planck collaboration's result \cite{Planck:2018vyg}. This is intriguing and could be related to the trend of $H_0$ decreasing with the increasing redshift, as it is observed in several works \cite{Wong:2019kwg,Krishnan:2020vaf,Dainotti:2021pqg,Dainotti:2022bzg,Colgain:2022rxy,Jia:2022ycc,Malekjani:2023dky,Vagnozzi:2023nrq}. We plan to analyze this trend in our model in a future work.

Inspecting Table \ref{tbd} we note the difference between the best fitted values of the matter density, $\Omega_m$, in the case of $\La$CDM model, and \emph{the effective matter density}, $\Omega_{m0} / b_3$, in the case of $\ga\ld$CDM model. Best fitted values of the effective matter density, $\Omega_{m0} / b_3$, are more consistent across the data sets compared to the $\Omega_m$ values in $\La$CDM. Likewise, \emph{the effective dark energy density}, $\Omega_{e0} / b_\gamma$, has consistent values across the data sets. We emphasize the important distinction between the dark energy density, $\Omega_{e0}$, and its effect on the expansion of the universe through the term, $\Omega_{e0} / b_\gamma$ as given in (\ref{H2}). Unlike in the case of the matter density and the effective matter density, which have approximate values with each other, the dark energy density and the \emph{effective} dark energy density could be quite different. This difference is especially pronounced in the best fit values of these quantities. We emphasize that the best fit values of the dark energy density ($\Omega_{e0}$) are found quite low, on the order of a few percent, very much different than the values inferred for the effective dark energy density ($\Omega_{e0} / b_\gamma$), which are approximate to the values found for $\Omega_{e0}$ in the $\La$CDM model. Thus this model empowers small amount of dark energy to have big impact on the expansion of the universe. One other interesting point worth emphasizing is the following: the \emph{effective} densities for the dark energy and the matter, $\Omega_{e0} / b_\gamma$ and $\Omega_{m0} / b_3$ respectively, have best fit values very close to the best fit values for dark energy and matter densities in the $\La$CDM model. Thus even though the \emph{true} densities could be quite different between the models, their effect on the expansion of the universe is almost the same in the $\ga\ld$CDM and the $\La$CDM models.

\begin{table*}[hbt!]
\centering
\def\arraystretch{1.1}
\begin{tabular}{|l|c|c|c|c|}
\hline 
\hline 
\textbf{Data set} & CC & CC+Pan+ & CC+BAO & CC+BAO+CMB \\
\hline 
\textbf{Model} & $\ga\ld$CDM & $\ga\ld$CDM & $\ga\ld$CDM & $\ga\ld$CDM \\
 & \textcolor{blue}{$\La$CDM} & \textcolor{blue}{$\La$CDM}
 & \textcolor{blue}{$\La$CDM} & \textcolor{blue}{$\La$CDM} \\
\hline
%\hline
$\mathbf{H_0}$
 & $69.37$ & $71.04$ & $67.78$ & $67.80$  \\
 & \textcolor{blue}{$69.43$} & \textcolor{blue}{$70.56$} & \textcolor{blue}{$71.13$} & \textcolor{blue}{$69.55$}   \\
%\hline
$\mathbf{\Omega_{m0}}$
 & $0.309$ & $0.271$ & $0.338$ & $0.326$ \\
 & \textcolor{blue}{$0.309$} & \textcolor{blue}{$0.325$} & \textcolor{blue}{$0.308$} & \textcolor{blue}{$0.285$}  \\
%\hline
$\log_{10}(\mathbf{\Omega_{s0}})$ &\; $< -10.45$\; &\; $< -11.12$\; &\; $< -15.84$\; &\; $< -11.74$\;  \\
%\hline
$\mathbf{\gamma}$
 & $0.015$ & $0.359$ & $0.023$ & $0.148$ \\
%\hline
$\mathbf{\delta}$
 & $0.007$ & $0.156$ & $0.011$ & $0.070$ \\
%\hline
\hline
$\mathbf{\Omega_{e0}}$
 & $0.042$ & $0.012$ & $0.023$ & $0.002$  \\
%\hline
$\mathbf{\Omega_{e0} / b_\gamma}$
 & $0.688$ & $0.667$ & $0.658$ & $0.645$ \\
%\hline
$\mathbf{\Omega_{m0} / b_3}$
 & $0.312$ & $0.333$ & $0.342$ & $0.355$ \\
\hline
\hline
\end{tabular} 
\caption{\rm Best fit values of the free ($H_0,\ \Omega_{m0},\ \Omega_{s0},\ \gamma$ and $\delta$) and some derived parameters ($\Omega_{e0}, \Omega_{e0} / b_\gamma$ and $\Omega_{m0} / b_3$) of the $\gamma\delta$CDM model for different data set combinations. For comparison we also provide best fit values of the parameters of the $\La$CDM model (for which $\delta,\ \gamma$ and $\Omega_{s0}$ all vanish) in blue.} 
\label{tbd}
\end{table*}

Accepting that the dark energy density, $\Omega_{e0}$, is just a few percent of the universe's energy bucket, is there an explanation for the origin of such a \emph{low} value for the density of dark energy? Interestingly, some recent works on the cosmological coupling of black holes \cite{Croker:2021duf,Farrah:2023opk} might provide an explanation. Hypothesis of these works is that a black hole's mass is coupled to cosmological expansion and thus it changes with redshift as
\be \label{BHM}
M (z) = \Big( \frac{1+z_i}{1+z} \Big)^k M (z_i) \qquad z \leqslant z_i\ ,
\ee
where $k$ is called cosmological coupling strength and black hole becomes cosmologically coupled at redshift $z_i$. 
It is also claimed in \cite{Croker:2020plg} that spatial distribution of these black holes as point sources becomes uniform on scales $\lesssim 200$ Mpc. Thus, as the universe expands the density of such a ``continuum fluid'' will decrease with a rate of $(1+z)^3$ \cite{Croker:2020plg}. Together with the black hole mass growth as given above in (\ref{BHM}) the contribution of this continuum fluid to the expansion of the universe will depend on the redshift with the power law $(1+z)^{(3-k)}$. The power $(3-k)$ is equivalent to $\gamma$ in our work (see equation (\ref{emt})). The best fit value of $(3-k)$, inferred by comparing supermassive black holes in five samples of elliptical galaxies at $z > 0.7$ to those in contemporary elliptical galaxies \cite{Farrah:2023opk} is well within the 1$\sigma$ regions of the $\gamma$ parameter for all combinations of data sets. Thus, our work is in agreement with the claims made in \cite{Croker:2021duf,Farrah:2023opk} and \cite{Croker:2020plg}. This whole line of argument is also in agreement with 
the theoretical considerations put forward long time ago by Gliner \cite{Gliner:1966}. Gliner argues that the gravitational collapse of a star do not end up in a singularity, but the force of gravitational contraction renders the pressure of matter to be negative, which effectively acts to contract the distribution of matter. Thus, the state of collapsed matter in a black hole interior can be considered as a source of dark energy.

%%%%%%%%%%.       Conclusions

\section{Conclusions \label{conc}} 

In this work, we investigated the cosmology of an ellipsoidal universe in the framework of $f(R)$ theory of gravity. Although we observe a homogeneous and isotropic universe at scales larger than roughly 100 $Mpc$, there are indications that this was possibly not always the case during the history of the universe. These are the observations that indicate possible anisotropy in the distribution of matter and the geometry of the universe as summarized in the introduction. In the standard cosmological paradigm, there are also so called ``tensions'' of the results from observations of early and late epochs of the universe. The most famous of these are the tensions in the value of the Hubble constant $H_0$ and the value of the growth parameter $S_8$ (see excellent reviews on cosmological tensions, such as \cite{Abdalla:2022yfr,Shah:2021onj,Freedman:2021ahq,Verde:2019ivm,DiValentino:2021izs,Perivolaropoulos:2021jda,Schoneberg:2021qvd,Knox:2019rjx}). There are many proposals in the literature to resolve these tensions, and the alternative gravity approach is one of them. In the present work we investigated the effect of perhaps the simplest extension of general relativity, namely $f(R)$ gravity on the evolution of the universe. There are many works that examines the evolution of various types of ellipsoidal universes in the general relativistic context, e.g., see \cite{Campanelli:2006vb,Cea:2022mtf,Campanelli:2007qn,Cea:2019gnu,Akarsu:2019pwn,Akarsu:2020pka,Akarsu:2021max,Tedesco:2018dbn,Amirhashchi:2018bic,Hossienkhani:2014zoa}. 

Our work is distinguished from other works on $f(R)$ gravity cosmology in the sense that we do not obtain the Friedmann equation in its general relativistic form. In the alternative gravity cosmologies, the usual approach is to treat the extra terms, which are additional to the general relativistic ones in the field equations, as an effective energy-momentum tensor and this way to obtain so called ``curvature'' contributions to the various energy densities. Here our approach is quite different. The Friedmann equation is not in the general relativistic form. The consequence of this modified form of the Friedmann equation (\ref{H2}) is enormous. The energy densities do not contribute the expansion of the universe as in the ordinary Friedmann equation. Depending of the value of the $\delta$ parameter (\ref{gamma}) the contribution of various energy densities to the Hubble parameter are enhanced or diminished. Existence of the anisotropic stress $\Omega_{s0}$ requires to have the $\delta$ parameter \cite{Leach2006} in the theory. Then not just the contribution of various energy densities to the modified Friedmann equation are weighted by the coefficients $b_n$ (\ref{bn}), but also their scale factor or redshift dependence shifts (\ref{H2}) by $a^\delta = (1+z)^{-\delta}$ compared to the standard formalism \cite{Mukhanov:2005sc}. Other than these changes, we find that this solution of the field equations do not allow a cosmological constant term as part of the $f(R)$ function. Nevertheless, one can consider the possibility of a dark energy term with redshift dependence $(1+z)^\gamma$, with $0<\gamma<2$ that originates from the matter Lagrangian. Together with the shift $(1+z)^{-\delta}$, the dark energy term in the modified Friedmann equation depends on the redshift by $(1+z)^{\gamma-\delta}$ (\ref{H2}).

We tested observational relevance of the new solution by best fitting to different data sets, such as the Pantheon type Ia supernovae (SNe Ia) data \cite{Pan-STARRS1:2017jku}, the cosmic chronometers (CC) Hubble data \cite{Jimenez:2001gg,Favale:2023lnp} and the Baryon Acoustic Oscillations (BAO) data \cite{Staicova:2021ntm} of the late universe, and the CMB data \cite{Planck:2018vyg} of the early universe.
We presented constraints, mean values with 68\% CL and the best fit values, on the free ($H_0,\ \Omega_{m0},\ \Omega_{s0},\ \gamma$ and $\delta$) and some derived parameters ($\Omega_{e0}, \Omega_{e0} / b_\gamma$ and $\Omega_{m0} / b_3$) of the $\gamma\delta$CDM model for different data set combinations as CC, CC+Pan+, CC+BAO and CC+BAO+CMB. We found that, in the $\gamma\delta$CDM model, the best fit values of the Hubble constant, $H_0$, the effective matter density, $\Omega_{m0} / b_3$, and the effective dark energy density $\Omega_{e0} / b_\gamma$ are in agreement with the best fit values, obtained in the $\La$CDM model, of the Hubble constant, $H_0$, the matter density, $\Omega_{m0}$, and the dark energy density $\Omega_{e0}$, respectively.

The important difference between the $\gamma\delta$CDM model and the $\La$CDM model manifests itself in the best fit value of the dark energy density, $\Omega_{e0}$, which, as explained, could be very different than the effective dark energy density, $\Omega_{e0} / b_\gamma$. We have observed that the best fit values for the dark energy density, $\Omega_{e0}$, are just a few percent of the universe's total energy. 
We noted that such a \emph{low} value for the density of dark energy could be explained with the cosmological coupling of black holes \cite{Croker:2021duf,Farrah:2023opk}. Hypothesis of these works is that a black hole's mass is coupled to cosmological expansion and its dependence on the scale factor is proportional to $a^k$, where $k$ is called cosmological coupling strength. As shown further in \cite{Croker:2020plg} the spatial distribution of these black holes as point sources becomes uniform on scales $\lesssim 200$ Mpc. Thus, as the universe expands the density of such a ``continuum fluid'' will decrease with a rate of $(1+z)^3$ \cite{Croker:2020plg}. Together with the black hole mass growth, the contribution of this continuum fluid to the expansion of the universe will depend on the scale factor with the power law $a^\gamma$, where $\gamma = 3-k$. The best fit value of $(3-k)$, inferred by comparing supermassive black holes in five samples of elliptical galaxies at $z > 0.7$ to those in contemporary elliptical galaxies \cite{Farrah:2023opk} is well within the 1$\sigma$ regions of the $\gamma$ parameter for all combinations of the data sets. Thus, our work is in agreement with the claims made in \cite{Croker:2021duf,Farrah:2023opk} and \cite{Croker:2020plg}. 
This whole line of argument has a theoretical basis put forward long time ago by Gliner \cite{Gliner:1966}, who argues that the state of collapsed matter in a black hole interior can be considered as a source of dark energy with negative pressure and $p+\rho \geqslant 0$. 

There are further cosmological tests to perform in order to check viability of our model. Firstly, we would like to test our model with further data sets \cite{Brout:2022vxf,Nunes:2020hzy,eBOSS:2020yzd,deCarvalho:2021azj}. With data sets we used in this work we obtained consistent best fit values for the Hubble constant across the data sets. This should be confirmed with data fit analysis with further data sets in order to claim that there is no tension in the value of the Hubble constant in our model. There is another important tension in the $\La$CDM model, which is the tension in the value of the growth parameter $S_8$ \cite{Abdalla:2022yfr,Perivolaropoulos:2021jda,DiValentino:2020vvd,Nunes:2021ipq}. We plan to analyze the value of the growth parameter in a future publication and check whether the tension can be resolved in our model. There are also tensions related with the ages of very old objects, such as quasars \cite{Jain:2005gu}, very old elliptic galaxies \cite{Vagnozzi:2021tjv}, and the very early galaxies recently observed by the James Webb Space Telescope \cite{Naidu:2022wia,Castellano:2022wia,Menci:2022wia,Labbe:2023wia}. We plan to calculate the ages of those very old objects in the universe from the redshift information and check whether our model better fits the age data than the $\La$CDM model and better explain the properties of the very early galaxies.

\section*{Acknowledgments} 

C.D. thanks \"{O}zg\"{u}r Akarsu, Nihan Kat\i rc\i , K. Yavuz Ek\c{s}i and Bar\i \c{s} Yap\i \c{s}kan for helpful discussions.
Vildan Kele\c{s} Tu\u{g}yano\u{g}lu is also supported by TUBITAK 2211/C Priority Areas of Domestic Ph.D. Scholar. The numerical calculations reported in this paper were partially performed at TUB\.{I}TAK ULAKB\.{I}M, High Performance and Grid Computing Center (TRUBA resources).

%\newpage

%%%%%%%%%%%%	References

\end{document}